**Modeling key characteristics of high-efficiency gallium arsenide solar cells**

A.V. Sachenko[1], V.P. Kostylyov[1,a], I.O. Sokolovskyi[1], and A.I. Shkrebtii[2]

[1]*V. Lashkaryov Institute of Semiconductor Physics, NAS of Ukraine, 03028 Kyiv, Ukraine*

[2]*Ontario Tech University, Oshawa, ON, L1G 0C5, Canada*

a) Author to whom correspondence should be addressed: vkostylyov@ukr.net

ABSTRACT

The paper proposes a theoretical approach to modeling the key characteristics of highly efficient gallium arsenide-based solar cells (SCs), using a one-dimensional SC model. The following recombination mechanisms are considered in the modeling: radiative recombination, interband Auger recombination, Shockley–Reed–Hall (SRH) recombination, surface recombination, recombination in the space-charge region (SCR), and recombination along the structure perimeter. A simple empirical formula is introduced to describe the recombination along the perimeter of the SC structure. The GaAs bandgap narrowing effect is also included. The main results are based on the assumption that the Shockley-Reed-Hall recombination time and the recombination time in the SCR are the same. The effect of photon recycling (re-emission and re-absorption) is accounted for in a model similar to the one we used previously to simulate key characteristics of high-efficiency monocrystalline silicon SCs. The model also uses absorption analysis at different gallium arsenide doping levels. Good agreement was achieved between the experiment and theory.

The results from this work can be used to optimize and enhance the characteristics of highly efficient SCs based on direct-bandgap semiconductors, especially gallium arsenide (GaAs).

## 1. Introduction

Gallium arsenide-based SCs currently demonstrate the highest photoconversion efficiency $\eta$ of 29.1% among single-junction SCs, compared with the very recent record of 27.8% for silicon SCs [1-3]. Accordingly, the record efficiency of gallium arsenide SC is closest to the Shockley-Queisser theoretical limit of 33.5% [1,2,4]. Further increasing the SC's efficiency significantly raises the requirements for the theoretical characterization of gallium arsenide-based devices. Understanding of the physics of photoconversion processes in GaAs-based SCs is also crucial for their widespread use in creating tandem and cascade SCs, which achieve efficiencies of 38.8% (5 junctions under AM1.5) and 47.6% (4 junctions under concentrated 665× AM1.5) [3].

Let's compare the physical descriptions of the photoconversion process and the modeling software for monocrystalline silicon and gallium arsenide solar cells. Currently, the theoretical framework and simulation programs for silicon-based devices are significantly more advanced and refined than those for gallium arsenide counterparts. For silicon, in addition to commercial professional high-level programs (Silvaco ATLAS, Synopsys Sentaurus TCAD), which allow modeling the characteristics of SCs (see, e.g., [5]), there are several specialized mid-level programs. Examples are Quokka [6], PC1D [7], PC3D [8], and others [9]. The computational software SCAPS-1D [10] and wxAMPS [11] are designed for numerical modeling of SCs and semiconductor heterostructures based on almost any semiconductor, including silicon. Given the nature of the approach we developed and described below, we are confident that developers and active users of such programs will find the material presented in this paper interesting and successfully apply for their SCs.

Until recently, analysis of current-flow mechanisms in GaAs SCs was carried out by considering, e.g., components with different coefficients of non-ideality, which did not, in particular, account for the effect of bandgap narrowing. At the same time, despite some differences between the direct bandgap GaAs and the indirect bandgap Si, both semiconductors share many similarities, including the fact that the same recombination mechanisms operate in both. In particular, for small-area SCs, the recombination rate along the device perimeter exhibits the same dependencies. However, the parameterization of recombination mechanisms for silicon SCs has recently been significantly improved, and several formulas have been proposed to describe Auger recombination in silicon, as analyzed in [12]. Furthermore, several authors have refined the value of the theoretical Shockley-Queisser limit for silicon SCs, and it is now known with high accuracy; no such refinement has yet been made for SCs based on gallium arsenide.

The key novelty of our approach to modeling efficiency in this work is that it accounts for all known current-flow mechanisms (except tunneling) and includes the effects of bandgap narrowing and recombination along the perimeter of the device structure. A distinctive feature of the proposed model is the analytical expression for the spectral dependence of the external quantum efficiency, $EQE(\lambda)$ near the absorption edge. The expression extends conventional formulations by accounting for deviations of light absorption from the ideal Lambertian limit through the introduction of the parameter b, which quantifies the degree of this deviation. Although this expression was originally developed and published for silicon solar cells [12], it is applied here for the first time to the modeling of GaAs-based solar cells. The proposed approach also enables optimization of the base thickness of high-efficiency GaAs solar cells. To the best of our knowledge, such a generalized, self-consistent methodology has not been reported in any prior published work.

As in [12], this study focuses exclusively on highly efficient SCs, in which the effective diffusion lengths of minority carriers significantly exceed the thickness of the base (absorber) regions.

The purpose of this work is to elevate the discussion, simulation, and parameter optimization of gallium arsenide-based SCs to the standard established for monocrystalline silicon SCs. Consequently, we anticipate that our work will engage not only researchers focused on gallium arsenide photoconversion, but also those specialists investigating similar phenomena in silicon, primarily concerning the calculation of critical solar cell parameters (see, for example, [12]).

## 2. Quantum efficiency and short-circuit current

External quantum efficiency $EQE(\lambda)$ allows the determination of the short-circuit current density $J_{SC}$ for incident radiation with the spectral density of the photon flux $I(\lambda)$ as

$$J_{SC} = q\int d\lambda \, EQE(\lambda)\, I(\lambda)\,, \tag{1}$$

where $q$ is the elementary charge. The magnitude of $EQE\ (\lambda)$ is determined by such factors as the chemical composition and morphology of the surface, the presence of an ITO coating or a grid for current collection, the absorption coefficient of the semiconductor, reflection, transmission, and re-absorption losses, as well as other factors.

A theoretical approach proposed in [13, 14] allows simulating $J_{SC}$ from known $EQE(\lambda)$ and the light reflection coefficient $R(\lambda)$ in the device structure. In [12], we proposed a simplified alternative approach that enables us to determine $J_{SC}$ and its dependence on the base thickness $d$, given only the known wavelength-dependent $EQE$. Its essence is as follows.

The analysis of experimental data shows that for silicon SCs, the wavelength-dependent $EQE(\lambda)$ obtained for samples of different thicknesses [15,16] can be divided into two regions. It is the short-wavelength region $\lambda_0 < \lambda < \lambda_1$, which we denote by an index $s$, where the external quantum yield $EQE_s(\lambda)$ practically does not depend on the thickness of the sample $d$, and the long-wavelength region $EQE_l(\lambda)$ with $\lambda > \lambda_1$, marked with an index $l$, where such a dependence is present. In the long-wavelength region near the absorption edge, a modified Lambertian of the following form is used to calculate the external quantum yield

$$EQE_l(\lambda, b) = \frac{f}{1 + b/\left[4\, n_r^2(\lambda)\, \alpha(\lambda) d\right]}, \tag{2}$$

where the fitting parameter $b$, which determines the shape of the $EQE_l(\lambda)$ curve, equals to the ratio of the average photon path length $4 n_r^2 d$ in a SC with ideal Lambertian scattering to the actual average photon path length $4 n_r^2 d / b$. Its physical meaning is that it determines the degree of deviation of the real absorption from the absorption with ideal Lambertian scattering, and the parameter $f$ is selected in such a way that the values $EQE_s(\lambda)$ and $EQE_l(\lambda)$ coincide at $\lambda = \lambda_1$. Here, $n_r(\lambda)$ is the refractive index of GaAs, and $\alpha(\lambda)$ is its absorption coefficient.

Note that when $b = f = 1$, expression (2) is equivalent to the formula for the ideal Lambertian absorption capability of SC [17].

In the region $\lambda < \lambda_1$, the experimental $EQE_s(\lambda)$ do not depend on the base thickness and are determined only by the losses due to reflection, shading, and absorption of light outside the base area of the SC.

Next, we will apply the proposed theoretical approach to describe high-efficiency gallium arsenide-based SCs, grown by the OMVPE method [18-20]. For these SCs, the short-circuit current density can be calculated for the AM 1.5 conditions by the formula

$$J_{SC}(d,b) = q\, f_1 \left[ \int_{\lambda_0}^{\lambda_1} I_{AM1,5}(\lambda) EQE(\lambda) d\lambda + f_2 \int_{\lambda_1}^{\lambda_d} I_{AM1,5}(\lambda) EQE(\lambda, b) d\lambda \right], \quad (3)$$

where $\lambda_0 = 300$ nm, $\lambda_1 = 800$ nm’, $\lambda_d = 900$ nm, $I_{AM1,5}(\lambda)$ is the spectral density of the photon flux under AM1.5 conditions, $EQE(\lambda)$ is the experimental external quantum efficiency in the $300 < \lambda < 800$ nm range, $EQE(\lambda, b)$ is determined by expression (2), the coefficient $f_1$ takes into account the absorption of light outside the boundaries of gallium arsenide SC. The value $f_2$ is chosen so that at $\lambda = 800$ nm $EQE(\lambda)$ and $EQE(\lambda, b)$ coincide.

Fig. 1 shows the experimental $EQE(\lambda)$, measured in gallium arsenide SCs in [18-20], and theoretical dependencies calculated using formula (2). In particular, for the SC from [18], the agreement of theory with experiment is achieved for the base thickness of 0.6 μm, and $b$=2. Theoretical dependencies of $EQE(\lambda)$ were also calculated using the expression for a flat structure with a fully reflecting rear surface [17]:

$$EQE(\lambda) = 1 - \exp(-2\alpha(\lambda) d). \quad (4)$$

The photon path length increase for the ideal Lambertian scattering (Eq. (2) at $b$ =1) compared to total reflection from the rear surface (Eq. (4)) is $2 n_r^2 = 25.6$.

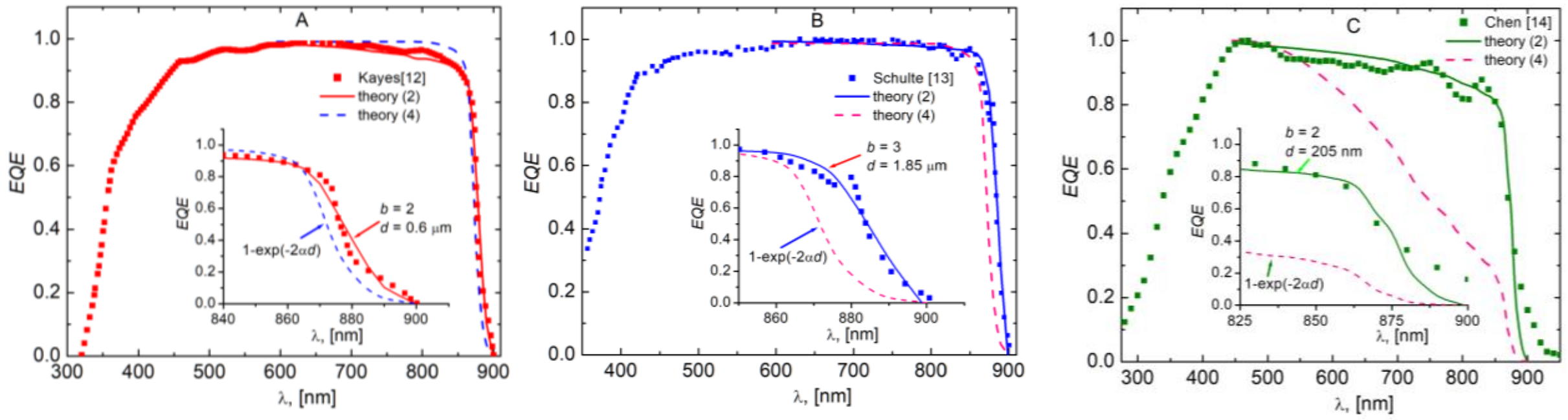

Fig. 1. Theoretical *EQE*(λ) dependencies obtained using Eq. (2) (solid lines) and Eq. (4) (dashed lines). The points show the experimental *EQE*(λ) for the SCs from [18-20]: A is from [18], B is from [19], and C is from [20], respectively.

The corresponding dependencies are also shown in Fig. 1, dashed lines. Spectral dependence of the gallium arsenide absorption coefficient α(λ*)* on the doping level from [21] was used in the calculations.

The coefficients *b* for GaAs SCs from [18-20], which we determined by fitting the theoretical dependencies to the experimental ones, are equal to 2, 3, and 2, respectively. As can be seen from Fig. 1, in the long-wave range, the experimental and theoretical dependencies obtained using Eq. (2) agree satisfactorily with each other. Some exceptions are shown in Fig. 1C: there is no agreement at *EQE*(λ) < 0.2, while the use of the theoretical expression (4) in this case does not agree with the experiment at all.

Let us focus on the physical meaning of external quantum efficiency in gallium arsenide-based SCs, described by Eq. (2). Unlike silicon SCs, in which similar dependencies exist only in textured structures, the situation is different in GaAs SCs.

Eq. (2) describes the external quantum efficiency *EQE*(λ) of solar cells in which light trapping occurs, i.e., due to multiple total internal reflections of rays into the sample as well as reabsorption/reemission photon cycles at the absorption edge, where the light absorption length 1/α(λ) exceeds the base (absorber) thickness. As the analysis of literature showed, the type (2) spectral dependencies of the external quantum efficiency in GaAs-based structures take place in asymmetric gallium arsenide structures with windows [18] and symmetric double AlGaAs-GaAs-AlGaAs heterostructures with non-textured surfaces, due to an average of at least 25 photon recycling events per photon (internal luminescence quantum efficiency of 99.7%; see Schnitzer [22]), in gallium arsenide-based SCs grown on acoustically spalled substrates with regions containing faceted surfaces with a relief of the order of half a micron [19], as well as in textured structures [20,24].

In [20,24], the applicability of the expression for $EQE(\lambda)$ given by Eq. (2) was demonstrated at the textured surfaces of the SC; in particular, in the case of [20], the silver back mirror was textured, and in the case of [24], the top surface of the $Al_{0.3}Ga_{0.7}As$ was chemically textured.

### 3. Concentration of intrinsic charge carriers in gallium arsenide

The intrinsic charge carrier concentration $n_i$ in gallium arsenide can be found as

$$n_i = \left[ N_C N_V \exp\left( - \frac{E_g(T)}{kT} \right) \right]^{1/2}, \tag{5}$$

where $N_C$ is the effective density of states in the conduction band, $N_V$ is the effective density of states in the valence band, $E_g$ is the bandgap, $T$ is the absolute temperature, and $k$ is the Boltzmann constant. Using the GaAs parameters from Johnson et al. [25], we get:

$$N_c N_v = 3.98 \cdot 10^{36}\ (\mathrm{T}/300)^3\ \mathrm{cm}^{-6}, \tag{6}$$

$$E_g = 1.519 - \frac{5.405 \cdot 10^{-4} T^2}{T + 204}\ [\mathrm{eV}] \tag{7}$$

and

$$n_i\ (298.15) \approx 1.78 \cdot 10^6\ \mathrm{cm}^{-3}. \tag{8}$$

Considering the effect of the gap narrowing in gallium arsenide, the intrinsic carrier concentration can be determined from

$$n_{ieff}(T, \Delta E_g) = n_i(T)\ \exp(\Delta E_g / 2kT). \tag{9}$$

The band narrowing depends on the doping level and the concentration of non-equilibrium carriers in GaAs, as considered in [26,27]. Fitting the experimental data, the authors of [27] proposed a formula for the band narrowing $\Delta E_g$:

$$\Delta E_g = A \cdot N^{1/3} + kT \cdot \ln\left[\Phi_{1/2}\left(E_F / kT\right)\right] - E_F, \tag{10}$$

where $A$=3.23·10$^{-8}$ eV·cm for $n$-type and $A$=2.55·10$^{-8}$ eV·cm for $p$-type GaAs, $\Phi_{1/2}$ is the Fermi-Dirac integral of order 1/2, and $E_F$ is the Fermi energy calculated, respectively, from the edge of the main carrier band (conduction band or valence band).

Fig. 2 shows the dependence of band narrowing on doping level when using Eq. (10), while the dots show adjusted experimental data from [27].

As mentioned above, the typical doping level of the base regions in gallium arsenide SCs is of the order of $10^{17}$ cm$^{-3}$. In some cases, a higher doping level was chosen (see [20]). Because the effective density of states for electrons in gallium arsenide is small, the electron degeneracy in the $n$-type occurs already at a doping level of $4 \cdot 10^{17}$ cm$^{-3}$. Therefore, the dependence of the band-narrowing on the doping level in this concentration range in *n*-type gallium arsenide becomes anomalous: the band-narrowing magnitude does not increase with increasing doping level; it actually decreases.

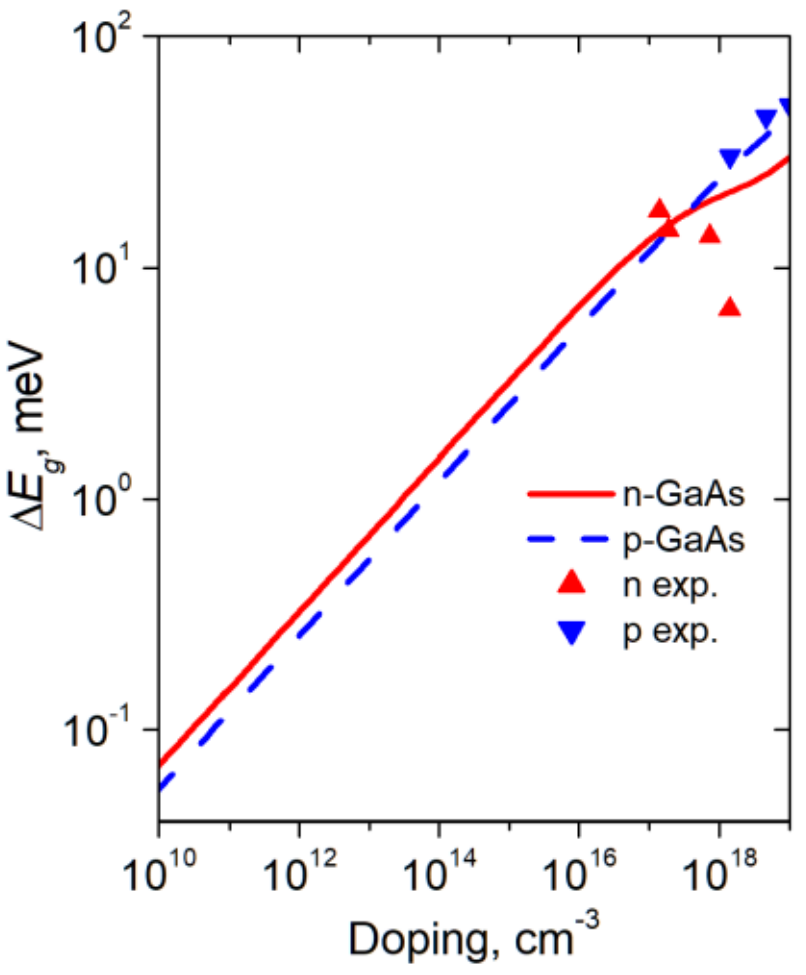

Fig. 2. Dependencies of the bandgap narrowing in GaAs $\Delta E_g$ on the doping level for *n*-type (red triangles are the experiment, red solid line is the theory) and *p*-type (blue inverted triangles are the experiment, blue dashed line is the theory).

Fig. 2 shows the dependence of the bandgap narrowing in *n*- and *p*-type gallium arsenide using data from [27] and the specified circumstances. Next, we used the given dependencies.

## 4. Lifetimes in gallium arsenide

The effective lifetime of minority charge carriers in gallium arsenide-based SCs, as in silicon SCs, is affected by intrinsic and extrinsic recombination mechanisms [22-24]:

$$\tau_{eff}^{-1} = \tau_{intr}^{-1} + \tau_{extr}^{-1}, \tag{11}$$

where

$$\tau_{intr} = \left(\tau_{rad}^{-1} + \tau_{Auger}^{-1}\right)^{-1} \tag{12}$$

is the lifetime for intrinsic (irremovable) recombination mechanisms, including radiative recombination and interband Auger recombination, $\tau_{extr}$ is the extrinsic lifetime due to bulk Shockley-Reed-Hall recombination $\tau_{SRH}$, surface recombination $\tau_{eff}^{s}$ and recombination in the space-charge region $\tau_{SCR}$. As shown in [14,25], recombination along the perimeter of the SC with lifetime $\tau_{per}$ also plays an important role in GaAs-based SCs. In general

$$\tau_{extr} = \left(\tau_{SRH}^{-1} + \left(\tau_{eff}^{s}\right)^{-1} + \tau_{SCR}^{-1} + \tau_{per}^{-1}\right)^{-1}. \tag{13}$$

It should be noted that in silicon, a notable contribution is also made by recombination mechanisms involving excitons, namely, radiative recombination of excitons and non-radiative recombination of excitons via the Auger mechanism via trap levels [12]. For its occurrence, the Bohr radius of the exciton must be small enough so that the hole and the electron are sufficiently close. In gallium arsenide, this mechanism does not work because the exciton Rydberg energy in GaAs is small (~4 meV), and the Bohr radius of the exciton is large.

The expression for the lifetime of radiative recombination in gallium arsenide can be obtained by transforming the corresponding expression for silicon given in [12] as follows:

$$\tau_{rad}^{-1} = B\,(1 - P_{RR})\,(n_0 + \Delta n), \tag{14}$$

where $B$ is the parameter of radiative recombination in gallium arsenide, $P_{RR}$ is the probability of photons re-absorption and re-emission, $n_0$ is the equilibrium concentration of electrons in the SC base, $\Delta n$ is the excess concentration of electron-hole pairs. $B$ can be found using the following relations:

$$B = \int_0^{\infty} dE\ B(E), \text{ where } \quad B(E) = \left( \frac{n_r(E)\,\alpha(E)\,E}{\pi\,c\,\hbar^{3/2}\,n_i} \right)^2 e^{-E/k\,T} \ . \tag{15}$$

Here, $n_r(E)$ is the refractive index, and $\alpha(E)$ is the absorption coefficient of gallium arsenide as a function of photon energy $E = hc/\lambda$ .

The probability of photons' re-absorption and re-emission is defined as

$$P_{RR} = B^{-1}\int_0^{\infty} dE\ A_{bb}(E)\,B(E), \tag{16}$$

and the magnitude of absorption $A_{bb}(E)$ is equal to

$$A_{bb}(E) = \frac{\alpha(E)}{\alpha(E) + \alpha_{np}(E) + b/(4n_r^2(E)d)}, \tag{17}$$

that is, the absorption is assumed not to be perfectly Lambertian. Expression (17) is similar to (2) and differs from the one given by Tiedje, Yablonovich, et al. [17] in that the coefficient $b$, greater than 1. Here, $\alpha_{np}$ is the non-photoactive component of absorption. To separate it from the photoactive one, we approximated the exciton peaks at the band-band absorption edge for each doping level.

Therefore, to obtain correct data for the radiative recombination parameter and the probability of photon re-absorption, it is necessary to construct the total and photoactive absorption coefficients in the material for an arbitrary conductivity type and doping level. This can be done using known absorption data [21] and modeling its edge, accounting for excitonic effects. Due to uncertainty in the absorption data, another approach is the shift of high-purity absorption with bandgap narrowing.

Dependencies of the effective coefficient of radiative recombination $B_{eff} = B\,(1 - P_{PR})$ on the effective thickness of the base $d_{eff} = d/b$ for GaAs are plotted in Fig. 3. As can be seen from the figure, the larger the effective thickness of the base, the smaller the value of $B_{eff}$ since this increases the probability of re-absorption of the emitted photon.

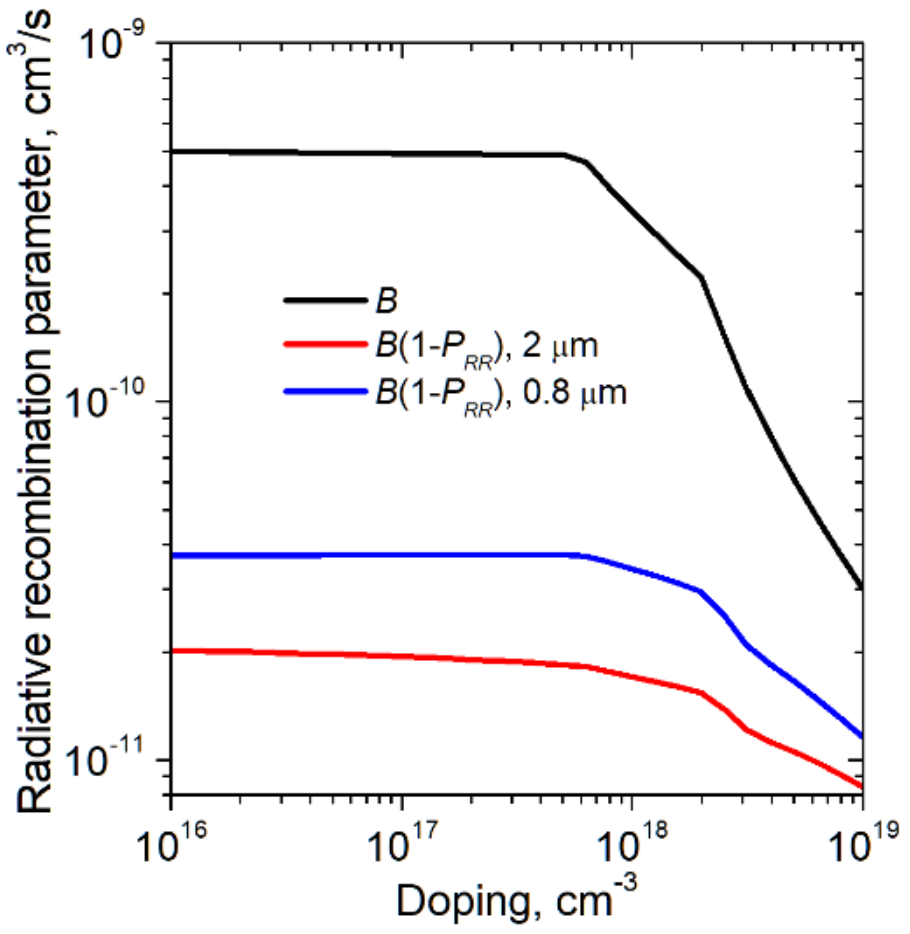

Fig. 3. Dependence of the effective radiative recombination coefficient in gallium arsenide $B_{eff}$, on the effective thickness $d_{eff}$.

For the Auger interband recombination lifetime, we used the expression

$$\tau_{Auger} = \frac{\Delta n}{C_e(n^2 p - n_0^2 p_0) + C_h(np^2 - n_0 p_0^2)}, \tag{18}$$

where $C_e$ and $C_h$ are the interband Auger recombination coefficients in gallium arsenide for electrons and holes, respectively, $n_0$, $n$, $p_0$, and $p$ are the equilibrium and non-equilibrium concentrations of electrons and holes. In the simulation, we used the Auger recombination constants $C_e = C_h = 7\cdot 10^{-30}$ cm$^6$ s$^{-1}$ [28].

The Schockley-Read-Hall lifetime, dependent on the doping and the excitation levels in the $n$-type GaAs, is described as follows:

$$\tau_{SRH} \cong \frac{(n_0 + n_1 + \Delta n) + b_r(p_1 + \Delta n)}{(n_0 + \Delta n) C_p N_t}, \tag{19}$$

where $b_r = C_p / C_n$, $C_p = V_p \sigma_p$, $C_n = V_n \sigma_n$, $V_p$ and $V_n$ are the average thermal velocities of holes and electrons, $\sigma_p$ and $\sigma_n$ are the cross-sections for hole and electron capture by the recombination level, $N_t$ is the recombination level concentration, $n_1 = n_i(T) e^{E_t/kT}$, $p_1 = n_i(T) e^{-E_t/kT}$ are the concentrations of electrons and holes when the Fermi level coincides with the energy of the recombination level, $E_t$ is the energy of deep levels, measured from the middle of the GaAs bandgap. Depending on the excess concentration of electron-hole pairs $\Delta n$, $\tau_{SRH}$ varies between low injection and high injection values. A similar expression holds for *p*- type GaAs.

The lifetime due to surface recombination $\tau_{eff}^{s}$ is defined as

$$\tau_{eff}^{s} = \left(\frac{S_S}{d}\right)^{-1}, \tag{20}$$

where $S_S$ is the total surface recombination velocity at the boundaries of the base region. Next, the $\Delta n$ dependencies of $S_S(\Delta n)$ are required to find the photoconversion efficiency and other key parameters. They can be found by solving the theoretical expression for $S_S(\Delta n)$ using a model of a discrete surface level located near the middle of the bandgap. In this case

$$S(\Delta n) = \frac{C_n C_p N_s (n_0 + \Delta n)}{C_n n_s + C_p p_s}. \tag{21}$$

Here $C_n = V_n \sigma_n$, $C_p = V_p \sigma_p$ is the electron and hole capture coefficients, respectively, $N_s$ is the surface level concentration, $n = n_0 + \Delta n$, and $p = p_0 + \Delta n$ are the non-equilibrium bulk electron and hole concentrations, respectively, $n_s = n\exp(y_s)$, $p_s = p\exp(-y_s)$ is the surface electron and hole concentrations, respectively, $y_s = \frac{q\varphi_s}{kT}$ is the dimensionless surface band bending.

Expression (21) can be used to find the dependencies $S_S(\Delta n)$ for both gallium arsenide-based SCs and silicon SCs. In the case of silicon, when a strong inversion of conductivity occurs (that is, the inequality $p_s >> n_0$ is satisfied) due to high concentrations of acceptors in the emitter at its boundary with the base region, the solution of equation (21) gives the dependence [12]

$$S_S(\Delta n) = S_{S0}\left(1 + \frac{\Delta n}{n_0}\right). \tag{22}$$

We will return to finding the $S_S(\Delta n)$ dependencies in the case of gallium arsenide later.

The lifetime $\tau_{SCR}$ due to recombination in SCR is defined as

$$\tau_{SCR} = \left(\frac{S_{SCR}}{d}\right)^{-1}, \tag{24}$$

where $S_{SCR}$ is the recombination rate in SCR.

The expression for the recombination rate in the SCR is obtained by integrating over the SCR layer the expression for the inverse Shockley-Reed-Hall lifetime (19) on the *x*-coordinate dependent electron and hole concentrations as a function of the band bending: $n(x) = (n_0 + \Delta n)e^{y(x)}$, $p(x) = (p_0 + \Delta p)e^{-y(x)}$.

$$S_{SCR} = \int_0^w 1/\tau_{SCR}(x)\ dx\ , \tag{25}$$

Substituting the expression for the Shockley–Read–Hall lifetime into (25), accounting for the above considerations, and writing the expressions for $n_1$ and $p_1$ in more detail, we obtain

$$S_{SCR}(\Delta n) = \int_0^w \frac{(n_0+\Delta n)dx}{\left[\left((n_0+\Delta n)e^{y(x)} + n_i(T)\exp\left(\frac{E_t}{kT}\right)\right) + b_r\left((p_0+\Delta n)e^{-y(x)} + n_i(T)\exp\left(-\frac{E_t}{kT}\right)\right)\right]\tau_R(x)}\ . \tag{26}$$

Here $\tau_R = 1/(C_p N_t^*)$ is the lifetime in the SCR, $N_t^*$ is the deep level concentration in SCR, $p_0$ is the equilibrium holes concentration in the bulk, $y(x)$ is the dimensionless electrostatic potential (bands bending) in SCR, $E_t$ is the energy of the deep level in the SCR in GaAs, measured from the middle of the bandgap, $n_i(T)$ is concentration of the intrinsic charge carriers, and $w$ is the SCR thickness:

$$w = \int_{y_0}^{y_w} \frac{L_D}{\left[\left(1+\frac{\Delta n}{n_0}\right)(e^y-1) - y + \frac{\Delta n}{n_0}(e^{-y}-1)\right]^{0.5}} dy\ , \tag{27}$$

where $y_0$ is the non-equilibrium dimensionless bands bending on the surface of a more weakly doped region (base), which depends on the level of excitation $\Delta n$ and is found from the integral neutrality equation, $y_w$ is the value of the dimensionless potential at the boundary of the SCR and the quasi-neutral volume, $L_D = (\varepsilon_0 \varepsilon_{\mathrm{GaAs}} kT / 2q^2 n_0)^{1/2}$ is the Debye length, $\varepsilon_0$ is the vacuum permittivity and $\varepsilon_{\mathrm{GaAs}}$ is the relative permittivity of gallium arsenide. For specific calculations, we used the values $y_w = -0.1$.

The expression for $w$ is obtained by double integration of the Poisson equation for the space charge region in the base.

If $\tau_R(x) = \text{const}$, the integral (26) can be reduced to the following. Going from integration over coordinate $x$ to integration over the dimensionless potential $y$, we obtain

$$S_{SCR}(\Delta n) = \int_{y_w}^{y0} \frac{(n_0+\Delta n)dy}{\left[\left((n_0+\Delta n)e^{y} + n_i(T)\exp\left(\frac{E_t-E_i}{kT}\right)\right) + b_r\left((p_0+\Delta n)e^{-y} + n_i(T)\exp\left(-\frac{E_t-E_i}{kT}\right)\right)\right]\tau_R} F\ , \tag{28}$$

where

$$\frac{dy}{dx} = \frac{1}{L_D}\left[(1+\Delta n/n_0)(e^y-1) - y + (p_0/n_0 + \Delta n/n_0)(e^{-y}-1)\right]^{1/2} = \frac{1}{F}\ , \tag{29}$$

$$F = \left(\frac{dy}{dx}\right)^{-1} = L_D\left[\left(1+\Delta n/n_0\right)(e^{y}-1) - y + \left(\frac{p_0}{n_0} + \Delta n/n_0\right)\left(e^{-y}-1\right)\right]^{-1/2} . \quad (30)$$

Expression (29) is the first integral of the Poisson equation.

In what follows, we will limit ourselves to the analysis of the case when the recombination level is near the middle of the bandgap. Consequently, terms in the denominator (26) proportional to $n_i(T)$, can be neglected.

To find the dependence of the non-equilibrium dimensionless potential on the coordinate $x$, it is necessary to use the solution of the Poisson equation (the second integral), which in quadrature has the following form:

$$x = \int_{y_0}^{y} \frac{L_D}{\left[\left(1+\frac{\Delta n}{n_0}\right)(e^{y_1}-1) - y_1 + \frac{\Delta n}{n_0}(e^{-y_1}-1)\right]^{0.5}} dy_1 . \quad (31)$$

The non-equilibrium dimensionless potential $y_0$ at $x$=0 is found from the solution of the integral electroneutrality equation in the form

$$N = \pm\left(\frac{2kT\varepsilon_0\varepsilon_{GaAs}}{q^2}\right)^{1/2}\left[\left(n_0+\Delta n\right)(e^{y_0}-1) - n_0 y_0 + \Delta n\left(e^{-y_0}-1\right)\right]^{1/2} , \quad (32)$$

where $qN$ is the surface charge density of acceptors in a $p$-$n$-junction, or in an anisotypic heterojunction.

In a more general form, the electroneutrality equation was given, in particular, in Appendix 1 of [29].

When the voltage applied to the junction increases, or the junction is illuminated, the potential $y_0$ decreases, and the SCR thickness $w$ decreases, i.e., a part of the SCR becomes neutral. In the region that has become neutral, recombination also occurs, with the rate equal to

$$S_{SCRn} = \left(\frac{w(\Delta n = 0) - w(\Delta n)}{\tau_R}\right)\frac{\left(n_0+\Delta n\right)}{\left(n_0+\Delta n\right)+b_R\Delta n} . \quad (33)$$

As the analysis shows, in high-efficiency silicon SCs [12], the $S_{SCRn}$ at excess carrier concentrations at the maximum power and open circuit is of the order of magnitude comparable to the $S_{SCR}$ value, since the lifetimes in SCR are much shorter than the lifetime in the neutral base. Therefore, when calculating the total recombination in the SCR $S_{SCRn}$, in the case of silicon, the following expression, which includes both components, should be used

$$S_{SCRs} = S_{SCR} + \left(\frac{w(\Delta n = 0) - w(\Delta n)}{\tau_R}\right)\frac{\left(n_0+\Delta n\right)}{\left(n_0+\Delta n\right)+b_R\Delta n} . \quad (34)$$

In gallium arsenide, the situation is different. As shown in our work [30], the doping level of the base material at the maximum photoconversion efficiency is about $10^{17}$ cm$^{-3}$. Therefore, it is logical to assume that in the SCR, the doping level is the same as in the base region, and the lifetime in the SCR is determined by the same deep level that determines the Shockley-Reed-Hall lifetime in the base, so that $\tau_{SRH} = \tau_R$ and the contribution from the part of the SCR that becomes neutral can be neglected. As can be seen from what follows, the use of the specified assumptions allows, in the first approximation, a good description of the experimental values of the key characteristics of SC based on gallium arsenide.

Next, we describe the algorithm for finding the experimental recombination rate in the SCR. Experimental value of $S_{SCR}^{\exp}$ is found using expressions

$$S_{SCR}^{\exp} = d\left[\left(\tau_{eff}^{\exp}\right)^{-1} - \left(\tau_{eff}^{k}\right)^{-1}\right]. \quad (35)$$

where $\tau_{eff}^{k}$ is the effective lifetime without the component describing the recombination time in the SCR:

$$\tau_{eff}^{k} = \left(\tau_{SRH}^{-1} + \tau_{eff}^{s\ -1} + \tau_{rad}^{-1} + \tau_{Auger}^{-1}\right)^{-1}. \quad (36)$$

**5. Expressions for calculating the characteristics of solar cells.**

Using the above expressions allows us to calculate the light voltage-current (I-V) characteristic, dark voltage-current characteristic, $J_{SC}(V_{OC})$, and the photoconversion efficiency.

The light voltage-current characteristic for SC is determined using the following relations:

$$I_L(V) = I_{SC} - I_r(V) - \frac{V + I_L R_s}{R_{SH}}, \quad (37)$$

$$I_r(V) = qA_{SC}\left(\frac{d}{\tau_{eff}^{b}} + S_{s0}\left(\frac{n_0}{n_p}\right)\left(1 + \frac{\Delta n}{n_0}\right) + S_{SCRs}\right)\Delta n(V), \quad (38)$$

$$\tau_{eff}^{b}(n) = \left[\frac{1}{\tau_{SRH}} + \frac{1}{\tau_{per}(n)} + \frac{1}{\tau_{rad}(n)} + \frac{1}{\tau_{Auger}(n)}\right]^{-1}, \quad (39)$$

$$\Delta n(V) = -\frac{n_0}{2} + \sqrt{\frac{n_0^2}{4} + n_i(T, \Delta E_g)^2\left(\exp\frac{q(V + I_L R_s)}{kT} - 1\right)}, \quad (40)$$

where $I_r(V)$ is the total current in the external circuit, $I_{Ph} \approx I_{SC}$ – is the photogeneration current, $I_{SC}$ is the short circuit current, $I_r(V)$ is the recombination (dark) current, $\tau_{eff}^{b}(n)$ is the bulk effective lifetime, $\tau_{per}(\Delta n)$ is the lifetime for recombination along the perimeter, $A_{SC}$ is the SC area, $V$ is the applied voltage, $R_s$ and $R_{SH}$ are the series and shunt resistance, $n = n_0 + \Delta n$ is the total concentration of electrons in the neutral volume of the base, and $\Delta E_g$ is the bandgap narrowing of the semiconductor. Equations (37) - (40) have a similar form when currents are replaced by their densities $I \rightarrow J = I/A_{SC}$.

From the expression (36), it is also possible to obtain the relationship for the open-circuit voltage, by setting the light current $I_L$=0, and $\Delta n(V) \equiv \Delta n_{OC}$. In this case, we have

$$V_{OC} = \frac{kT}{q} \ln\left(1 + \frac{\Delta n_{OC}\left(n_0 + \Delta n_{OC}\right)}{n_i(T)^2 \exp(\Delta E_g / kT)}\right). \tag{41}$$

Multiplying the current $I_L(V)$ by the applied voltage $V$, we get power $P(V)$, and from the maximum condition $dP/dV$ we find the voltage at the point of maximum power selection, $V_m$. Substituting $V_m$ in Eq. (37), we get the current at the point of maximum power $I_m$. This allows the photoconversion efficiency to be calculated in the usual way $\eta$ and the fill factor of the light I-V characteristic $FF$:

$$\eta = \frac{J_m V_m}{P_s}, \tag{42}$$

where $P_s$ is the radiation density, incident on the SC surface under AM1.5. conditions.

$$FF = \frac{J_m V_m}{J_{SC} V_{OC}}. \tag{43}$$

Expression for the dark current $I_D(V)$ has the following form:

$$I_D(V) = \frac{qA_{SC}\Delta nd}{\tau_{eff}(\Delta n)} + \frac{V - I_D R_s}{R_{SH}}. \tag{44}$$

$$\Delta n(V) = -\frac{n_0}{2} + \sqrt{\frac{n_0^2}{4} + n_i(T, \Delta E_g)^2\left(\exp\frac{q\left(V - I_D R_s\right)}{kT} - 1\right)}. \tag{45}$$

Finally, the dependencies of the short circuit current on the open circuit voltage are determined from equation (37). From the condition $I_L(V_{OC}) = 0$, we get

$$I_{SC}(V_{OC}) = \frac{qA_{SC}\Delta nd}{\tau_{eff}(\Delta n)} - \frac{V_{OC}}{R_{SH}}, \tag{46}$$

$$\Delta n(V_{OC}) = -\frac{n_0}{2} + \sqrt{\frac{n_0^2}{4} + n_i(T, \Delta E_g)^2\left(\exp\frac{qV_{OC}}{kT} - 1\right)}. \tag{47}$$

Note that the series resistance $R_s$ does not affect the $I_{SC}(V_{OC})$ dependencies.

The effective lifetime as a function of the excess concentration can be obtained both from the open-circuit voltage-dependent short-circuit current and the dark current:

$$\tau_{eff} = qd\left(\frac{\Delta n(V_{OC})}{J_{SC}(V_{OC})}\right), \tag{48}$$

$$\tau_{eff} = qd\left(\frac{\Delta n(V)}{J_D(V)}\right) \tag{49}$$

If $J_{SC}(V_{OC})$ dependencies are measured, then $\Delta n(V_{OC}) \equiv \Delta n(n_0, T, V_{OC})$, the procedure for obtaining both experimental and theoretical $\tau_{eff}(\Delta n)$ is sufficiently simple. When the dark current $\Delta n(V) \equiv \Delta n(n_0, T, R_s, V, J_D(V))$ is measured, to find $\tau_{eff}(\Delta n)$ one has to use the dependencies $\Delta n(n_0, T, R_s, V, J_D(V))$.

## 6. Experimental results and their comparison with theory

### 6.1. Analysis of dark currents and discussion of the results

To correctly calculate the excess concentration-dependent recombination rate in the SCR for the SCs from [18-20], it is necessary to find the surface charge density $N$ for each case. First, we evaluate the potential barriers for the SCs from [19] and [20]. The total barrier height can be found from Eq. 29:

$$\varphi = kT\ln\left(\frac{n_n p_p}{n_i^2}\right), \tag{50}$$

where $n_n$ and $p_p$ are the concentrations of charge carriers in the base and the emitter.

The barrier in the lightly doped region is higher than in the heavily doped region and can be found as:

$$\varphi_L = \varphi\frac{p_p}{p_p + n_n}. \tag{51}$$

For the SC from [19], the base doping level is $2.5\cdot10^{17}$ cm$^{-3}$, and the emitter doping level ranges from $10^{16}$ to $5\cdot10^{18}$ cm$^{-3}$. For certainty, we consider the emitter doping to be $2\cdot10^{18}$ cm$^{-3}$. Then the total barrier height is 1.38 V, and the potential drop in the base is 1.23 V. For the doping level of $2.5\cdot10^{17}$ cm$^{-3}$, the $N$ value is equal to $2.05\cdot10^{12}$ cm$^{-2}$.

For the SC from [20], the doping levels in the base and emitter are the same and equal to $10^{18}$ cm$^{-3}$, and the barrier value is 1.40 V. Since in [20] the p-n junction is symmetric, the potential drop in the base is 0.7 V. Using the integral neutrality formula, we can verify that such a barrier is realized at $N \approx 3\ 10^{12}$cm$^{-2}$.

The most difficult part is determining the $N$ value for the SC from [18]. Assuming a base charge carrier concentration of the SC is $3\cdot10^{17}$ cm$^{-3}$, and an emitter charge carrier concentration of $\geq 2\cdot10^{18}$ cm$^{-3}$, it follows that $N\geq 2\cdot10^{12}$ cm$^{-2}$. This assumption is justified by our calculations of the photoconversion efficiency and dark current for this SC across various base- and emitter-carrier concentrations. Using these specific concentrations and the same calculation parameters, we achieved agreement between theoretical and experimental results for both dark and light I-V characteristics.

As is known, the maximum contribution to the recombination in the SCR comes from a thin layer in the depletion region, localized in the vicinity of the layer where the minority carrier concentration equals the bulk majority carrier concentration, that is, the condition

$$\varphi_I = kT \ln\left(\frac{n_n}{n_i}\right) \qquad (52)$$

has to be satisfied. Here $\varphi_I$ represents the potential at the boundary of the depletion layer in the base region and the emitter, at which $p(\varphi_I) = n_n$. If the actual potential at the boundary of the base region with the emitter exceeds the specified $\varphi_I$ value, the recombination rate in the SCR reaches its maximum and remains constant despite any further increases in potential.

For the SCs from [18] and [19], $\varphi_I$ is close to 0.66 V, while the actual potential at the boundary of the base region with the emitter is much higher. Therefore, when $N$=$2\cdot10^{12}$ cm$^{-2}$, the conditions for saturation of the recombination rate, after reaching its maximum, are well satisfied. Furthermore, when using for calculations $N$=$3\cdot10^{12}$ or $N = 4\cdot10^{12}$ cm$^{-2}$, the recombination rate in the SCR will not change. For the SCs from [20], $\varphi_i$ =0.7 V, which is equal to the potential at the boundary of the base region with the emitter. As the calculations show, in the case when the ratio $C_p / C_n < 1$, the saturation of the recombination rate in the SCR occurs slightly earlier.

Let us obtain the dependencies of the effective recombination rate on the excess concentration of charge carriers for SCs from [18-20]. Since the SCs under investigation have sufficiently small areas, another recombination type is present, namely, recombination along the SC perimeter. Its contribution has been analyzed in several works, in particular, by Belghachi [31]. As will be shown below, agreement between theory and experiment is achieved if we use the following empirical formula for the recombination rate along the perimeter

$$J_{per} = qS_{per}\Delta n\,,\ S_{per} = S_{0per}\left(\frac{\Delta n + n_i^2 / n_0}{n_1}\right)^{-r}, \qquad (53)$$

where $S_{0per}$ is the rate of recombination along the perimeter at low excitation levels, $0 < r < 1$, $n_1$ is the normalization concentration.

In this case, it is possible to determine the experimental sum of the recombination rates in the SCR and along the perimeter. This can be done using the following formulas

$$S_{sum}^{\exp} = d\left(\left(\tau_{eff}^{\exp}\right)^{-1} - \left(\tau_{eff}^{k}\right)^{-1}\right). \qquad (54)$$

where $\tau_{eff}^{k}$ is the effective lifetime without components describing the recombination times in the SCR and along the SC perimeter. It equals:

$$\tau_{eff}^{k} = \left(\tau_{SRH}^{-1} + \tau_{eff}^{s\ -1} + \tau_{rad}^{-1} + \tau_{Auger}^{-1}\right)^{-1}. \qquad (55)$$

The theoretical value $S_{sum}^{theor}$ is determined from the expression

$$S_{sum}^{theor} = S_{SC} + S_{per}. \qquad (56)$$

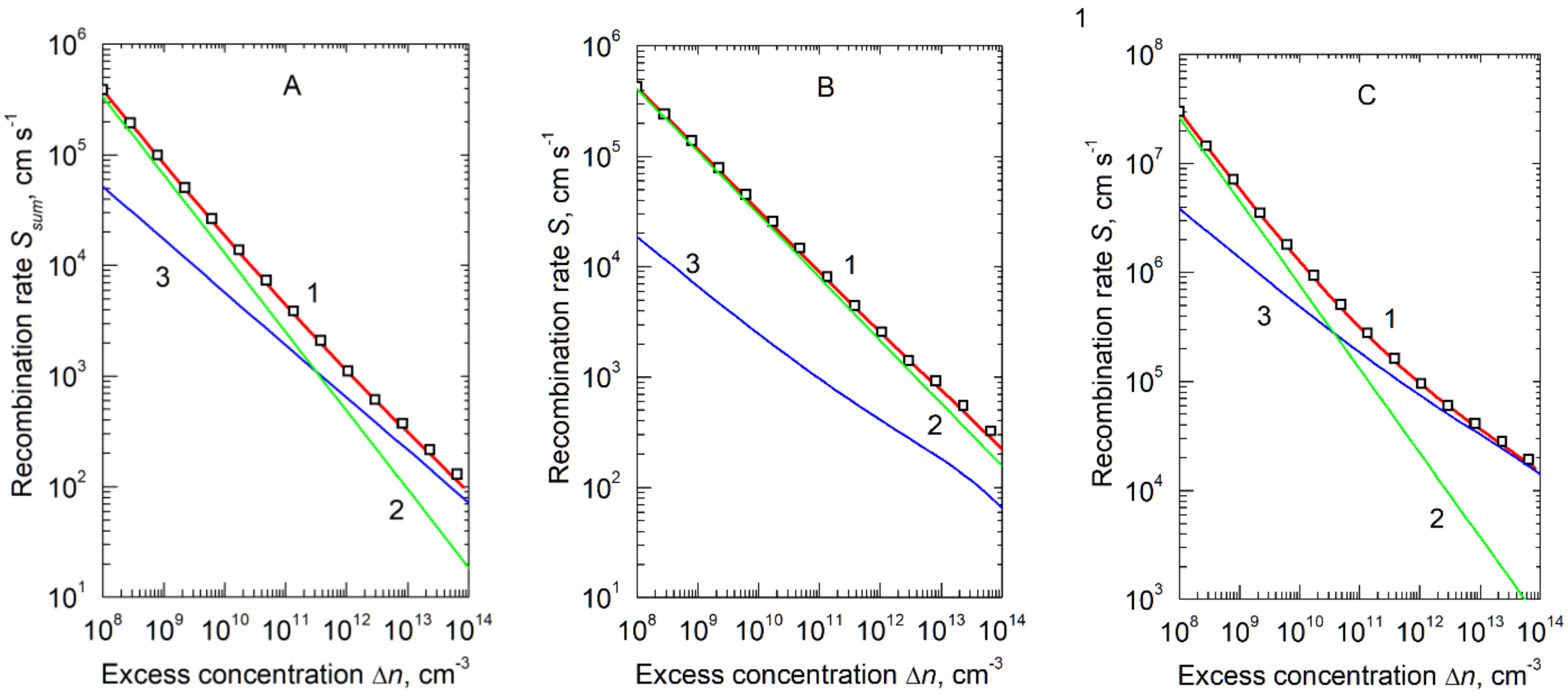

Fig. 4. Experimental total recombination rates in the SCR and along the SC perimeter $S_{sum}^{\exp}$ (squares) on the excess concentration, and the lines describe theoretical dependencies $S_{sum}^{theor}$ (lines 1) consistent with them for SCs from [18-20] (A - C, respectively). Lines 2 show the theoretical recombination velocity along the perimeter, and lines 3 show the theoretical recombination velocity in the SCR.

Fig. 4A-4C compares experimental $S_{sum}^{\exp}$ and corresponding theoretical $S_{sum}^{theor}$ dependencies for the SCs, studied in [18-20]. The parameters required to achieve agreement between the experimental and theoretical total recombination rates are given in Tables 1 and 2. These parameters were used to plot the current and subsequent figures.

Table 1. Solar cell parameters used to fit the voltage-dependent experimental and theoretical dark currents.

| SC | $n_0$,cm$^{-3}$ | $\tau_{SRH}$, s | $S_0$,cm/s | $d$, μm | $b$ | $R_S$, Ohm·cm$^2$ | $R_{SH}$,Ohm·cm$^2$ | $b_r$ |
|---|---|---|---|---|---|---|---|---|
| 1 | $3\cdot10^{17}$ | $8\cdot10^{-6}$ | 400 | 0.6 | 2 | 1.66 | $1.7\cdot10^{6}$ | $10^{-3}$ |
| 2 | $2.5\cdot10^{17}$ | $1.1\cdot10^{-7}$ | $2\cdot10^{3}$ | 1.85 | 3 | 1.38 | $1.1\cdot10^{7}$ | $10^{3}$ |
| 3 | $1\cdot10^{18}$ | $5.65\ 10^{-9}$ | $4\cdot10^{4}$ | 0.103 | 2 | 0.87 | $4.5\cdot10^{3}$ | $10^{3}$ |

Table 2. Parameters used to match the voltage-dependent experimental and theoretical dark currents.

| SC | $f$ | $B_{eff}$. cm$^3$/s | $S_{0\,per}$, cm/s | $n_1$,cm$^{-3}$ | $r$ |
|---|---|---|---|---|---|
| 1 | 1.36 | $6.64\cdot10^{-11}$ | $2.7\cdot10^{3}$ | $10^{11}$ | 0.71 |
| 2 | 1.38 | $6.98\cdot10^{-11}$ | $8\cdot10^{3}$ | $10^{11}$ | 0.57 |
| 3 | 1.61 | $2.58\cdot10^{-10}$ | $1.4\cdot10^{5}$ | $10^{11}$ | 0.76 |

As can be seen from Table 1, the Shockley-Read-Hall lifetimes in the structure with an n-type base are much longer than in the SC with a p-type base. This result correlates with the data on the Shockley-Read-Hall lifetimes in SC based on gallium arsenide, summarized in the work of Lumb [32]. Such behavior is typical for gallium arsenide [33].

Tables 1 and 2 give the parameters used to match the experimental and theoretical dark current dependencies on the voltage (see Fig. 4, A–C), with the notation “1” is for the SCs studied in [18], SCs 2 from [19], and SCs 3 from [20].

Next, we plot the voltage-dependent experimental and theoretical dark I-V characteristics from [18-20]. Figure 5 shows the experimental dark I-V characteristics, measured in [18-20], for GaAs SCs and the calculated values that agree with them. Calculations were carried out for AM1.5 illumination conditions at 25°C. In the calculations, the intrinsic carrier concentration in gallium arsenide $n_i$ was assumed to be $1.78\cdot10^{6}$ cm$^{-3}$ (8).

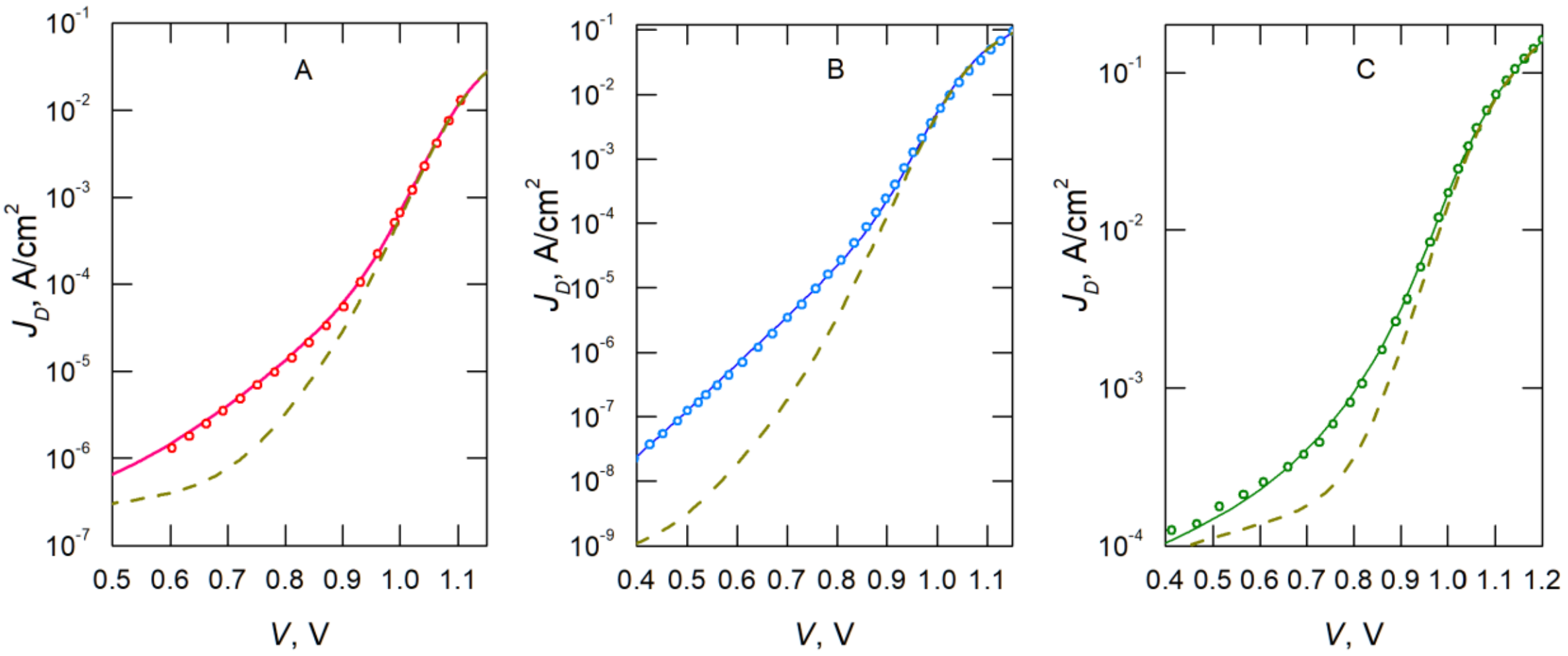

Fig. 5. Experimental (points) and theoretical (lines) dependencies of dark currents on the applied voltage for SCs from works [18-20]: A is for SCs from work [18], B is from [19], and C is from work [20]. Dashed lines correspond to theoretical dark currents without accounting for the recombination along the SC perimeter.

We initially plotted Fig. 5A-5C, considering only Shockley–Reed–Hall recombination, surface recombination, recombination in the SCR, radiative recombination, and Auger recombination (dashed lines). As can be seen from the above graphs, good agreement between the theory and experiment can only be achieved if the recombination along the SC perimeter is included.

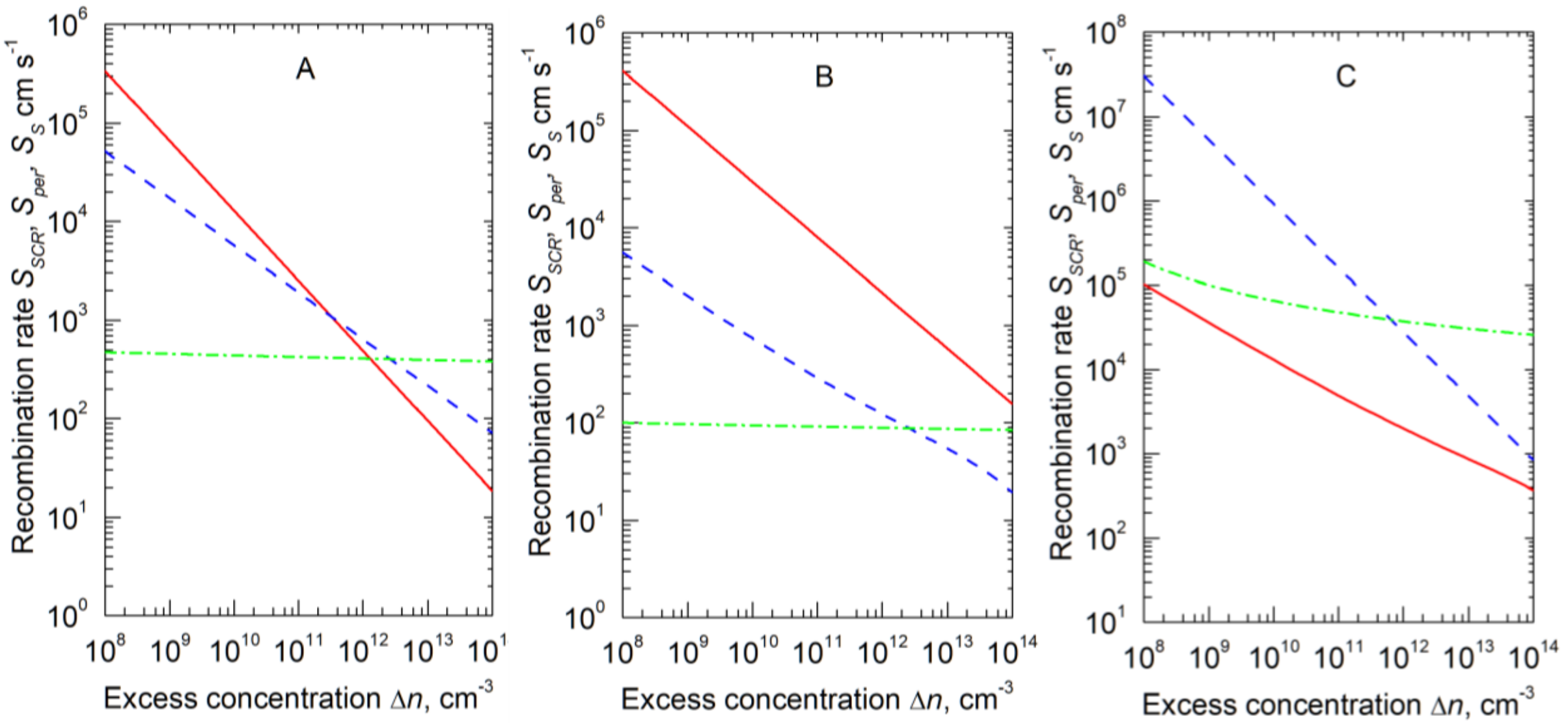

Fig. 6. Recombination in the SCR (red lines), along the SC perimeter (blue dashed lines), and surface recombination (green, dot-dashed lines) versus the excess charge carrier concentration $\Delta n$ for the solar cells from [18-20], respectively.

As can be seen from Fig. 6, the recombination rates in the SCR decrease with increasing $\Delta n$ more slowly than the recombination rates along the perimeter. Therefore, in the case of SC [18] and SC [20], the recombination rate along the perimeter dominates at small values of $\Delta n$, while at large values $\Delta n$, the recombination rate in the SCR dominates. The opposite situation is realized in SC [19]. In it, the recombination rate along the perimeter is much higher than the recombination rate in the SCR in the entire range of $\Delta n$ changes and therefore practically determines the total recombination rate.

The dependencies of the surface recombination rate $S$ on the excess concentration were found from the solution of equation (21) together with the integral neutrality equation (32). Calculations showed that the nature of these dependencies is determined by the surface charge density $qN$ at the junction boundary. Since in the case of SC [18-20] these densities are $2 \cdot 10^{12}$, $2 \cdot 10^{12}$ and $3 \cdot 10^{12}$ $cm^{-2}$, dependencies $S(\Delta n)$ at small and moderate values of $\Delta n$, not exceeding $10^{14}$ $cm^{-3}$, have a decreasing character. However, the decrease occurs much more slowly than the change in the value of $\Delta n$. In this case, the surface recombination rate is determined by its value at the maximum power point. Table 1 shows these $S_0$ values for SC [18-20].

All SCs fabricated in [18–20] were grown by the OMVPE method. It should be noted that the best agreement between the experimental and theoretical dark current curves occurs when the ratio of the hole-to-electron capture cross sections is 0.001. This value correlates with the literature on traps in gallium arsenide grown by OMVPE and HVPE; in [34-37], the dominant role of the EL2 trap was shown in this case. This trap is associated with the As antisite, with an energy position of 0.82 eV below the conduction band and a ratio of hole and electron capture cross sections $\sigma_p / \sigma_n = 0.001$. In the work of Schulte et al. [34], when modeling GaAs-based SCs grown by the OMVPE method at atmospheric pressure, the dominant defect was considered to be an electron trap EL2.

The electron diffusion coefficient in p-type GaAs at a doping level of $10^{17}$ $cm^{-3}$ is 90 $cm^2$ $s^{-1}$ [38,39]. The use of the recombination parameters given in Tables 1 and 2 gives for SC [18] effective electron lifetimes $(3\text{-}5) \cdot 10^{-8}$ s, which corresponds to diffusion lengths of the order of 15-20 micrometers. Even a decrease in the electron diffusion coefficient at higher doping levels does not significantly change the situation. For an n-type semiconductor, the hole diffusion coefficient at a doping level of $10^{17}$ $cm^{-3}$ is equal to 5 $cm^2$ $s^{-1}$. Using the same effective hole lifetimes as in the n-type, the diffusion lengths in this case are on the order of 4–5 micrometers. Therefore, even in this instance, the diffusion length exceeds the thickness of the base region by a factor of 5–20. A similar situation occurs for SCs [19] and [20]. Therefore, because the diffusion lengths are significantly larger than the base region thicknesses in the GaAs SCs under consideration, the requirements for the minority carriers' effective diffusion lengths are well satisfied.

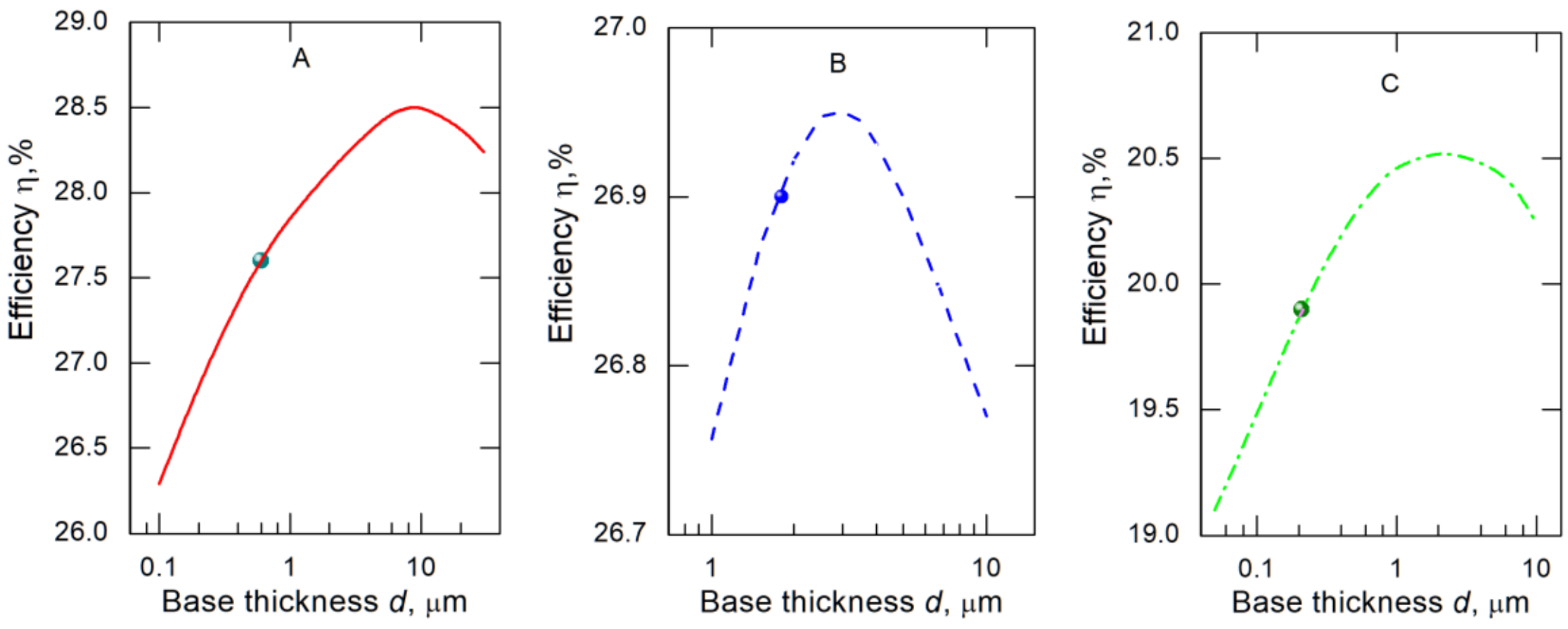

Fig. 7. Theoretical photoconversion efficiencies $\eta$ on the thickness of the base region $d$ for the SCs from [18] (A), [19] (B) and [20] (C). The calculations used the parameters given in Tables 1 and 2.

Fig. 7a-7c show the dependence of the photoconversion efficiency in SCs [18-20] on the base thickness. All three curves presented in the figure demonstrate a maximum. Their increasing sections are associated with an increase in the short-circuit current and the dominance of radiative recombination, the rate of which decreases with increasing base thickness. The decrease occurs when Shockley-Read-Hall recombination dominates. Since the Shockley-Read-Hall lifetimes in SCs with a p-type base are small, the maximum photoconversion efficiency is realized in them at significantly smaller base thicknesses than in SCs with an n-type base. The obtained dependencies allow optimizing the base thickness. As follows from Fig. 7b, the maximum photoconversion efficiency in SCs [19] occurs at a base thickness of about 2 μm, i.e., the base thickness in this SC is close to the optimal one that we calculated. In the case of SC [20], the maximum photoconversion efficiency is also found at a thickness of the order of 2 μm, which is significantly different from the experimental value. However, the peak efficiency of 20.53% is not significantly different from the experimental value for this SC. The most interesting situation is realized in SC [18]. Since the Shockley-Reed-Hall lifetime in this SC is very large (of the order of $10^{-5}$ s), its maximum photoconversion efficiency occurs at a base thickness of 7 μm. In addition, the photoconversion efficiency at the specified thickness reaches 28.94%, significantly exceeding the initial value of 27.6%.

In summary, the proposed method calculates the short-circuit current to determine the theoretical efficiency of GaAs SCs as a function of the base thickness, thereby facilitating optimization of the base region.

## 6.2. Analysis of photoconversion efficiencies and effective lifetimes

Fig. 8 shows the experimental light I-V characteristics and output power $P(V)$ *vs.* applied voltage for the GaAs-based solar cells from [18-20], along with the corresponding calculated values.

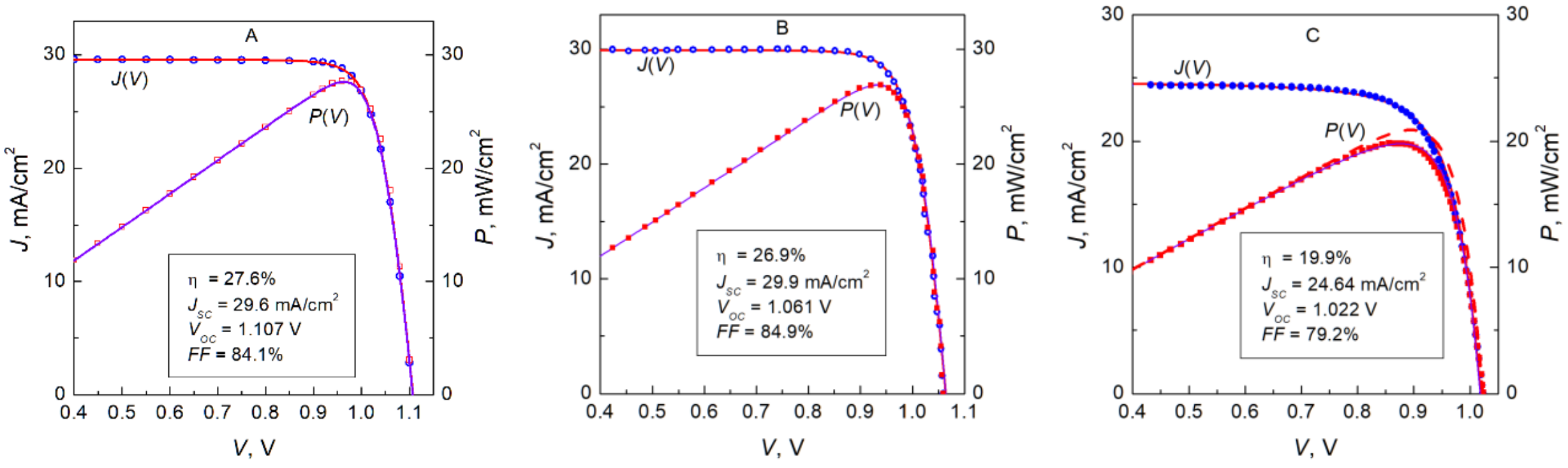

Fig. 8. Experimental (points) and theoretical (lines) dependencies of light $J(V)$ characteristics, and output power on the applied voltage $P(V)$ for SCs from [18-20]: A from [18], B from [19], and C from [20], respectively. In Fig. 8C, the dashed line shows the dependence of $P(V)$ ignoring the recombination along the SC perimeter. The calculated dependencies with and without perimeter recombination in Figs. 8A and 8B visually coincide.

The agreement between experiment and theory is good. Some of the differences in dark and light parameters may be because in [19,20] dark and light currents were measured on different structures with similar parameters, whereas in [18] they were measured on the same structure.

It should be noted that the recombination along the perimeter, while strongly influencing the dark I-V characteristics at low and medium applied voltages, has a weak effect on the photoconversion efficiency. Furthermore, there is no correlation between the change in dark currents and the photoconversion efficiency. Thus, for SC [18-20], the decrease in photoconversion efficiency due to recombination along the perimeter is 0.09, 0.28 and 1.2 %, respectively, which in relative units is 0.34, 1 and 6 %. That is, the largest efficiency decrease occurs for SC [20]. At the same time, the largest change in the dark current due to the recombination along the perimeter occurs for SC [19].

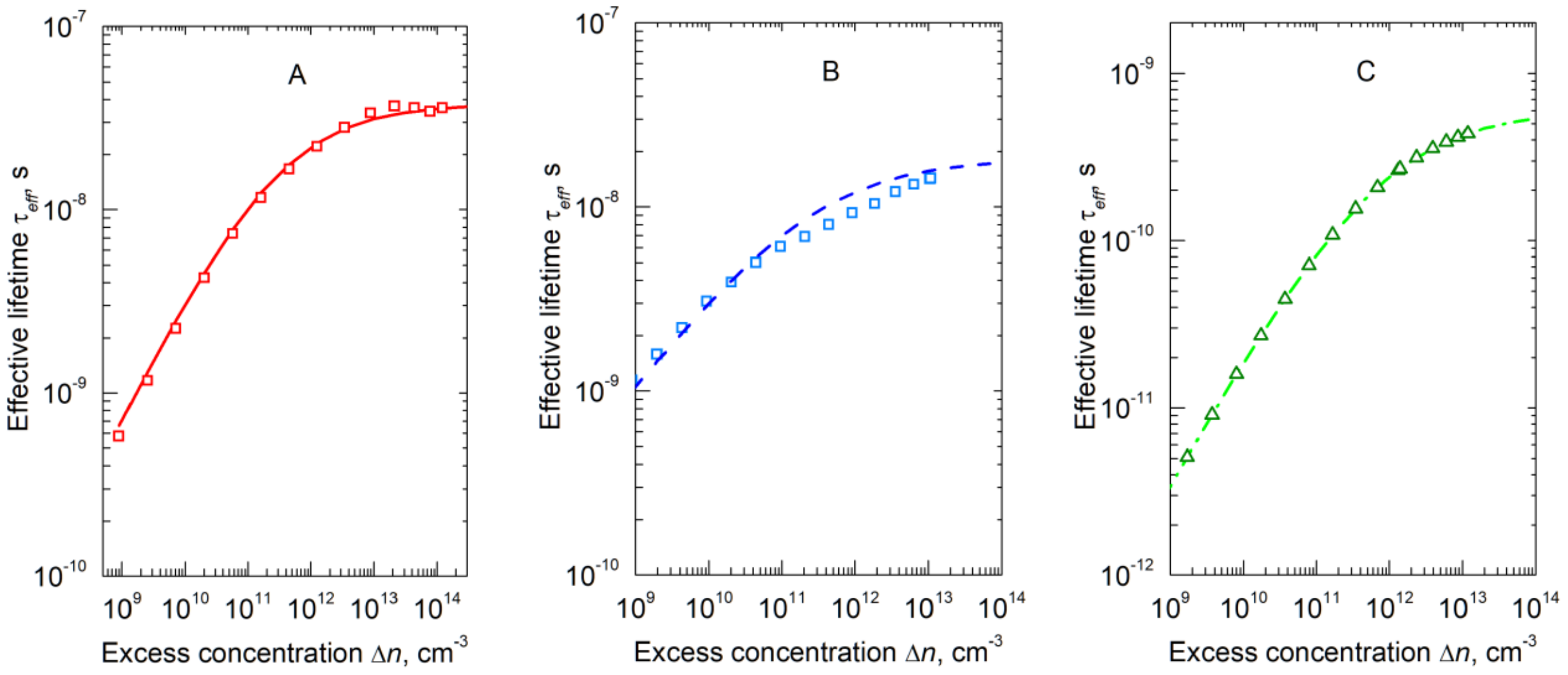

Fig. 9. Theoretical (lines) and experimental (points) dependencies of the effective lifetime on excess concentration, obtained for SCs from [18-20] (A-C, respectively).

Fig 9. shows the theoretical and experimental effective lifetimes as a function of excess concentration for SC from [18-20]. The experimental lifetime was obtained from expressions (48-53), and the theoretical lifetime was calculated using formulas

$$\tau_{eff} = \left(\tau_{rad}^{-1} + \tau_{Auger}^{-1} + \tau_{s}^{-1} + \tau_{SRH}^{-1} + \tau_{sc}^{-1} + \tau_{Per}^{-1}\right)^{-1}, \quad (57)$$

where

$$\tau_{rad} = \left((n_0 + \Delta n)B_{eff}\right)^{-1}, \quad (58)$$

$$\tau_{Auger} = \left((n_0 + \Delta n)^2 C_{Auger}\right)^{-1}, \quad (59)$$

$$\tau_{s} = \left((n_0 + \Delta n)\,\frac{S_0}{dn_0}\right)^{-1}, \quad (60)$$

$$\tau_{R} = \tau_{SRH}, \quad (61)$$

$$\tau_{SCR} = \left((n_0 + \Delta n)\,\frac{S_{SCR}}{dn_0}\right)^{-1}, \quad (62)$$

$$\tau_{Per} = \left((n_0 + \Delta n)\,\frac{S_{per}}{dn_0}\right)^{-1}. \quad (63)$$

As shown in Fig. 9, the effective lifetimes increase, tending toward saturation, and the theoretical and experimental values coincide. The obtained lifetime values are larger for SCs from [18,19] than those from [20], but all of them are much smaller than in the case of silicon. On the one hand, this is due to significantly lower Shockley-Reed-Hall recombination times in gallium arsenide, and a significant effect of recombination along the perimeter, on the other hand.

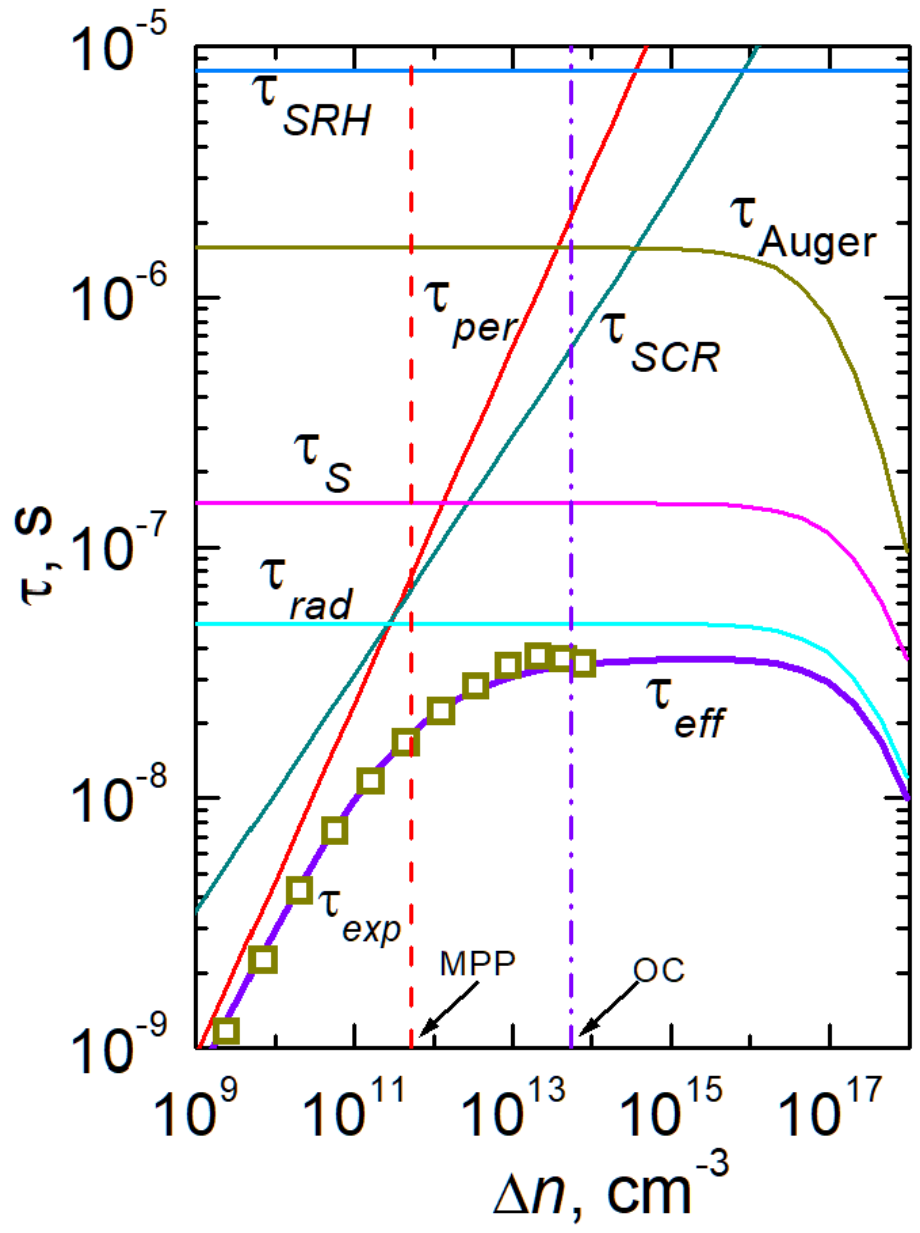

Fig. 10. Experimental (squares) and theoretical (solid lines) effective lifetime $\tau_{eff}$ on the excess charge carrier concentration $\Delta n$ for SC from [18] and its contributing parts: Shockley-Reed-Hall recombination $\tau_{SRH}$, Auger recombination $\tau_{Auger}$, radiative recombination $\tau_{rad}$, recombination in SCR $\tau_{SCR}$, recombination along the SC perimeter $\tau_{per}$, and surface recombination $\tau_S$.

Fig. 10 shows the theoretical dependencies for the effective lifetime and its components, as described by Eqs. (54) - (60), and using the parameters that we obtained for the SC from [18]. It can be seen that the dependence of the effective lifetime on the excess concentration of charge carriers $\Delta n$ in this SC at small $\Delta n$ is determined by recombination in the SCR and recombination along the SC perimeter. In contrast, at medium and large $\Delta n$, it is determined mainly by radiative recombination.

Similar results were obtained for the effective lifetime components for SCs from [19,20], except that for SC from [19] surface recombination plays a larger role, and for SC from [20] it dominates.

The results of calculations of the effective lifetime and its components for the SC from [18] were also used by us to analyze the distribution of recombination losses in this SC by the recombination mechanisms – Shockley-Reed-Hall, $\tau_{SRH}$, Auger, $\tau_{Auger}$, radiative recombination $\tau_{rad}$, recombination in the SCR $\tau_{SC}$, recombination along the SC perimeter $\tau_{per}$, surface recombination $\tau_S$. This analysis was performed in two operating modes: maximum power and open-circuit. All results are summarized in Fig. 11.

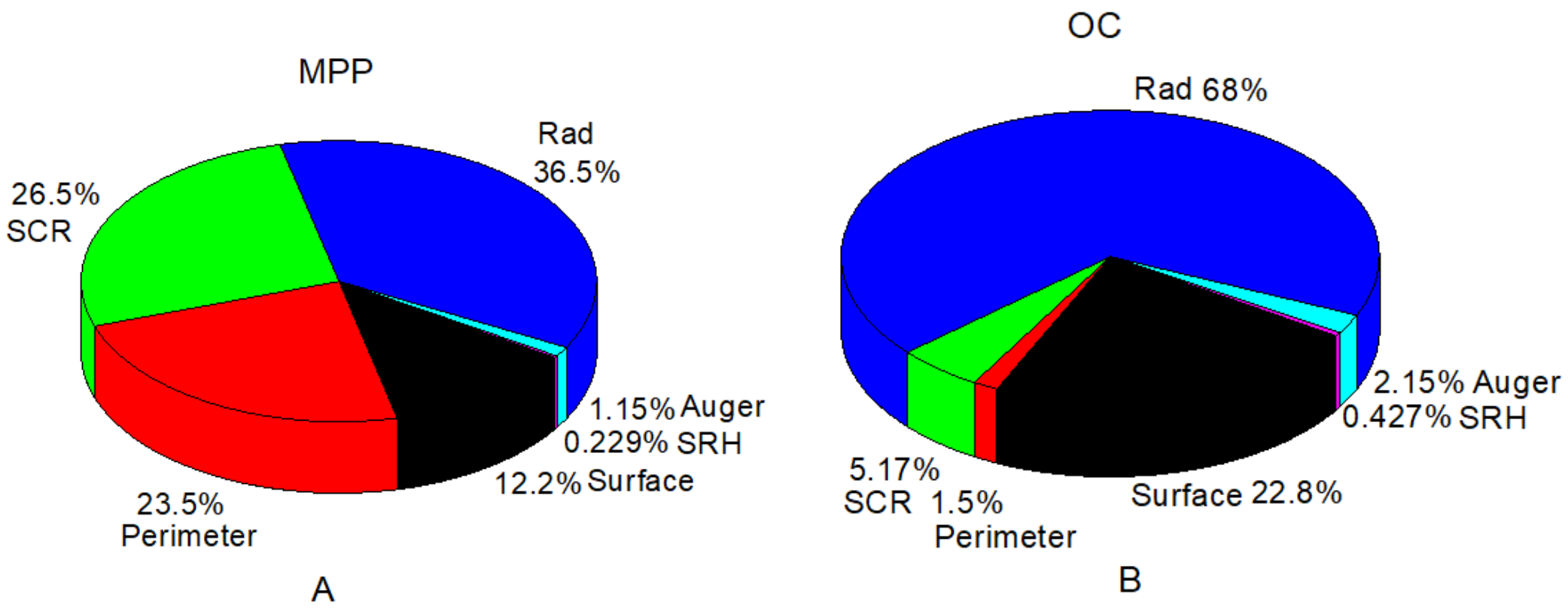

Fig. 11. Distribution of losses by recombination mechanisms at the maximum power point (A) (MPP, $V$=$V_{MPP}$) and at the open circuit point (B) at $V$=$V_{OC}$ in SC [18].

The data presented indicate that at the point of maximum power, radiative recombination dominates, highlighting the high perfection of the SC (referencing Yablonovich [5]: “A great solar cell also needs to be a great light-emitting diode”). Recombination in the SCR and recombination along the SC perimeter also make a significant contribution. At the open-circuit voltage, almost all recombination occurs through the radiation channel, which confirms the high perfection of this SC. Surface recombination makes a significant contribution, SRH recombination and Auger recombination make an order of magnitude smaller contribution, which explains the high open-circuit voltage of this SC.

It is worth noting that while previous works [18–20,33,40–42] addressed these recombination mechanisms in gallium arsenide in certain combinations, none considered them simultaneously.

## 7. Conclusions

In conclusion, this article introduces a theoretical approach that enhances the understanding of the physics of photoconversion processes. The novelty of the approach to modeling efficiency in this work is that it accounts for all known current transport mechanisms (except tunneling), and also includes the effects of band narrowing, photon recycling and recombination along the perimeter of the device structure. In addition, the proposed approach is self-consistent. It enables accurate comparison of calculated and experimental dependencies for the effective lifetime as a function of the excess concentration $\tau_{eff}(\Delta n)$, dark current-voltage and light current-voltage characteristics, as well as voltage-dependent output power in highly efficient direct-gap structures, particularly gallium-arsenide photovoltaic devices, where the diffusion lengths in the base regions significantly exceed their thickness. This work is based on the assumption that the spectral dependence of $EQE(\lambda)$ in the long-wave absorption region can be described by the empirical expression $EQE(\lambda,d) = 1/(1+b/4n_r^2(\lambda)\alpha(\lambda)d)$, where $b$ is an empirical coefficient greater than 1. The agreement between the theoretical dependencies of $EQE(\lambda)$ and the experimental data for the SCs from [18-20] was obtained when using the coefficient $b$ equal to 2, 3 and 2, respectively. Using this formula, the base thickness of highly efficient GaAs-based SCS was optimized. Furthermore, the proposed approach enabled the calculation of the theoretical dependencies of the effective radiative recombination coefficient $B_{eff}$.

This study proposes an empirical expression for the recombination current along the perimeter of the SC. We performed theoretical calculations of the dark and illuminated current-voltage characteristics, as well as the dependence of output power on voltage under the assumption that the Shockley-Reed-Hall lifetimes in the base regions and the recombination times in the SCR are equal. The good agreement between theory and experiment indicates the validity of the assumptions and the derived SC parameters. Ultimately, this approach allows for optimization of high-efficiency SCs based on semiconductors with a direct bandgap, such as gallium arsenide, indium phosphide, perovskites, CIGS, and others.

**ACKNOWLEDGMENTS**

This work has been supported by the National Academy of Sciences of Ukraine through the budget Program No. III-4-21 of fundamental research.

The data that support the findings of this study are available from the corresponding author upon reasonable request.

**Declaration of competing interest**

The authors declare that they have no known competing financial interests or personal relationships that could have influenced the work reported in this paper.

**Author Contributions**

**A. V. Sachenko**: Conceptualization (equal); Formal analysis (equal); Methodology (equal); Investigation (equal); Validation (equal); Writing original draft (equal). **V. P. Kostylyov**: Conceptualization (equal); Formal analysis (equal); Validation (equal); Visualization; Writing– review & editing (equal); Funding acquisition. **I. O. Sokolovskyi**: Data curation (equal); Investigation (equal); Software. **A. I. Shkrebtii**: Data curation (equal); Resources (equal); Writing– review & editing (equal).